\documentclass[aps,twocolumn]{revtex4}

\usepackage{amsmath}
\usepackage{graphicx}
\usepackage{color}

\newcommand{\bra}[1]{\langle#1|}
\newcommand{\ket}[1]{|#1\rangle}
\renewcommand{\vec}[1]{{\bf{#1}}}

\begin{document}

\title{Full counting statistics of heteronuclear molecules from Feshbach-assisted photoassociation}
\author{A. Nunnenkamp, D. Meiser, and P. Meystre}
\affiliation{Department of Physics, University of Arizona, Tucson, AZ 85721, USA}

\begin{abstract}
We study the effects of quantum statistics on the counting statistics of ultracold heteronuclear molecules formed by Feshbach-assisted photoassociation [Phys.~Rev.~Lett.~{\bf 93}, 140405 (2004)]. Exploiting the formal similarities with sum frequency generation and using quantum optics methods we consider the cases where the molecules are formed from atoms out of two Bose-Einstein condensates, out of a Bose-Einstein condensate and a gas of degenerate fermions, and out of two degenerate Fermi gases with and without superfluidity. Bosons are treated in a single mode approximation and fermions in a degenerate model. In these approximations we can numerically solve the master equations describing the system's dynamics and thus we find the full counting statistics of the molecular modes. The full quantum dynamics calculations are complemented by mean field calculations and short time perturbative expansions. While the molecule production rates are very similar in all three cases at this level of approximation, differences show up in the counting statistics of the molecular fields. The intermediate field of closed-channel molecules is for short times second-order coherent if the molecules are formed from two Bose-Einstein condensates or a Bose-Fermi mixture. They show counting statistics similar to a thermal field if formed from two normal Fermi gases. The coherence properties of molecule formation in two superfluid Fermi gases are intermediate between the two previous cases. In all cases the final field of deeply-bound molecules is found to be twice as noisy as that of the intermediate state. This is a consequence of its coupling to the lossy optical cavity in our model, which acts as an input port for quantum noise, much like the situation in an optical beam splitter.
\end{abstract}

\maketitle

\section{Introduction}

The ability to create molecules from ultracold atoms by means of magnetic Feshbach resonances \cite{Donley02, Durr04} or photoassociation \cite{Wynar00} has opened up exciting new areas of research from cold collisions physics to the study of strongly correlated systems \cite{Zoller00, Carr05, Miyakawa05} and the BEC-BCS cross-over \cite{Leggett80, Holland01, Regal04, Zwierlein05, Bartenstein05, Chen05}.
Both fermionic and bosonic atoms have been successfully converted into molecules whose condensation has also been achieved \cite{Greiner03, Zwierlein03, Jochim03}.
More recently, Feshbach resonances between different atomic species \cite{Stan04, Inouye04} have been observed, leading to the potential production of heteronuclear molecules.

In this context the question arises how the quantum statistics of the composing atoms influence the molecule formation process. Both Wouters {\it et al.}~\cite{Wouters03} and Dannenberg {\it et al.}~\cite{Dannenberg03} have studied the association of fermionic molecules from a boson-fermion mixture. They found that atom-molecule oscillations dominate the dynamics, similar to the case of two bosonic species.

In this paper we analyze molecule formation in quantum gas mixtures of atoms of all possible combinations of statistics. We find that at the level of molecule \emph{numbers} the molecule formation dynamics looks very similar in all three cases and that differences due to the different quantum statistics only show up in higher order correlation functions. We are therefore led to calculate the full counting statistics of the molecular fields which is a computationally hard problem, but contains complete information about the molecule number and all its moments.

Measurements of higher order correlations are typically difficult so that, in spite of their value in characterizing matter wave fields, progress in that direction has been made only recently. \"Ottl {\it et al.}~\cite{Ottl05} have determined the full counting statistics of an atom laser beam by single-atom detection within a high-finesse cavity. Following a proposal by Altman {\it et al.}~\cite{Altman04}, noise correlation experiments in the spatial domain have been used to measure the second-order coherence function of pair-correlated atoms from molecular dissociation \cite{Greiner05} and from an optical lattice in the Mott insulator regime \cite{Folling05}.

We focus on molecule production in a two-step process consisting of both a magnetic Feshbach resonance and photoassociation \cite{Kokkelmans01, Mackie02, Search04}. The idea is to use a two-photon Raman process to transfer the closed-channel molecules -- to which the free atoms are coupled through the hyperfine interaction -- to a much more stable deeply-bound state, where they are not so prone to three-body losses. This method is similar to the recent experiment by Partridge {\it et al.} \cite{Partridge05}, in which optical spectroscopy was used to probe the two-body state of paired $^6$Li atoms near a Feshbach resonance. One difference, however, is that we are considering here the Raman transfer inside an optical cavity that allows to select the final vibrational state of the molecules.

We investigate the cases where the molecules are produced from atomic condensates, from a mixture of bosons and fermions, as well as from fermions. In all cases the atomic sample is assumed to be at temperature $T=0$, and in the case of fermions we consider both the case of a normal Fermi gas and of a superfluid system. We derive master equations that describe the molecule formation process in the presence of cavity losses. For small particle numbers, these equations can be solved numerically in the single mode approximation for bosons and the degenerate approximation for fermions. Thus, we find the full counting statistics of the molecules \cite{Meiser05}.

We compare these numerical results to a perturbative analysis of the correlation functions and to mean-field calculations valid for short times. They show good agreement with the full numerical solutions of the master equations in the limit of large particle numbers.

We find that the qualitative features of the molecule production \emph{rates} are similar in all cases. As has been previously demonstrated e.g.~for the case of four-wave mixing \cite{Moore01,Ketterle01}, this is because the formation of superradiant collective states in fermions with Fermi energies $E_F$ at times short compared to the dephasing time $h/E_F$ leads to similar effects as the Bose enhancement for bosons.

Significant differences are, however, evident in the full counting \emph{statistics} presented here.
Atomic condensates and Bose-Fermi mixtures lead to coherent molecule statistics in the intermediate state, whereas normal Fermi gases lead to molecule statistics similar to those of a thermal field. The coherence properties of molecules formed from two superfluid Fermi gases is intermediate between the two.

The second-order coherence of the deeply-bound mole\-cules is particularly interesting, as this field is twice as noisy as that of the intermediate state. The additional noise finds its origin in the vacuum fluctuations of the cavity field, a situation somewhat analogous to that of an optical beam splitter, where vacuum noise is injected through the empty input port.

The remainder of this paper is organized as follows. Section II discusses the case of molecule formation from condensed bosonic atoms and shows in particular that the optical cavity acts as an input port for quantum noise that increases the quantum fluctuations of the molecular field. Section III addresses the case of Feshbach-assisted photoassociation from a Bose-Fermi mixture that consists of a BEC and a degenerate Fermi gas and shows that under appropriate circumstances that system can be mapped to the previous situation of two heteronuclear BECs. In section IV we consider the case of two degenerate Fermi gases with and without superfluidity. We conclude the paper with a discussion of our results in section V.

\section{\label{twobecs}Two Bose-Einstein condensates}

\subsection{Model}

We consider first the Feshbach-assisted photoassociation of condensates of bosonic atoms of species $A$ and $B$ into molecules. The atoms are prepared in a two component Bose-Einstein condensate. A sudden change of the magnetic field strength projects the scattering state of two free atoms onto the tunable molecular state. The ratio of open to closed channel admixture of this mole\-cular state determines atom-mole\-cule coupling and size of the molecular state. In entrance-channel dominated Fesh\-bach resonances a strong coupling between open and closed channel leads to molecular states which are predominately in the open channel and thus have a substantial overlap with the wavefunctions of free atomic pairs. This explains the large coherent atom-molecule oscillations such as in the Ramsey-type experiments \cite{Donley02}. In narrow resonances outside the universal regime the molecular state is mostly in the closed channel which reduces its size compared to the very weakly-bound molecules in broad resonances but has some open channel admixture which ensures that the molecular state is not orthogonal to the free atom scattering states. For a detailed discussion of molecular production in Feshbach resonances we refer to \cite{Goral04, Kohler06}. In this paper we assume that the molecular state can be approximately described by a quantum field satisfying bosonic commutation relations. This procedure is valid since the density $n_0$ and the molecular size i.e.~range of the molecular wavefunction $r_0$ satisfy $n_0r_0^3\ll 1$.

\begin{figure}
\includegraphics{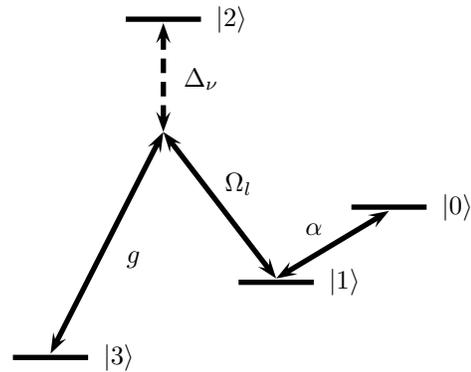}
\caption{Schematic of the atomic and molecular levels. The free atom pairs in the relative hyperfine state $\ket{0}$ can be taken from two BECs, a BEC and a Fermi gas or two Fermi gases, the latter with or without superfluid order parameter. Accordingly, the molecules in state $\ket{1}$, $\ket{2}$ and $\ket{3}$ will be described by a single mode, if they are bosons, or by many degenerate modes, if they are fermions.}
\label{levelscheme}
\end{figure}

The free atom pairs with total spin configuration $\ket{0}$ are coupled to closed-channel molecules with spin state $\ket{1}$. These closed-channel molecules are then coupled to a manifold of electronically excited molecular states $\ket{2_\nu}$ by a classical light field of frequency $\omega_l$ and Rabi frequency $\Omega_\nu(\vec R)$, where $\nu$ labels the vibrational levels. These states are finally coupled via a cavity mode $u(\vec R)$ of frequency $\omega_c$ to the electro-vibrational molecular ground state $\ket{3}$ with coupling strength $g$. Regarding the strength of the coupling between level $\ket{1}$ and $\ket{3}$ we restrict ourselves to the case $\kappa \gg g \Omega_{\nu} / \Delta_{\nu}$, where $\kappa$ is the linewidth of the cavity and the detuning $\Delta_{\nu}$ is defined below, i.e. we are not in the strong coupling limit of cavity quantum electrodynamics. The atomic and molecular levels together with the couplings between them are illustrated in Fig. \ref{levelscheme}.

At zero temperature and for sufficiently weak atom-atom and molecule-molecule interactions, excitations and quantum depletion can be neglected. Here we limit ourselves to this case and some implications of atom-atom and molecule-molecule interactions are discussed below. Then the center-of-mass motion of atom pairs and molecules can be treated in a single-mode approximation with center-of-mass wave functions $\phi_i(\vec R)$ and corresponding center-of-mass energies $\epsilon_i$. 

Denoting the internal energy of the molecular state $\ket{i}$ relative to that of the unbound atom pairs $\ket{0}$ by $\omega_i$, $i=1,2,3$, we assume that the detuning $\Delta_\nu = (\omega_{2,\nu} - \omega_1) - \omega_l \approx (\omega_{2,\nu} - \omega_3) - \omega_c \gg |g|,|\Omega_l|$ so that we can safely adiabatically eliminate the upper states $\ket{2_\nu}$, leading to effective two-photon Raman transitions between the states $\ket{1}$ and $\ket{3}$. The mean field shifts of each level have been absorbed in the energies $\omega_i$ but in all that follows we will neglect the change in the mean field energies due to the changing populations of the various modes.
In an interaction picture, in which the state $\ket{3}$ has the energy of the two-photon detuning $\delta =
(\omega_3 + \epsilon_3)-(\omega_1 + \epsilon_1)-(\omega_l - \omega_c)$ and $\omega = \omega_1 + (\epsilon_1 - \epsilon_A - \epsilon_B)$ equals the binding energy of the Feshbach molecule, Feshbach-assisted photoassociation is then described by the effective Hamiltonian \cite{Winkler05, Pazy05, Kheruntsyan05, Jack06}

\begin{equation}
\hat H = \delta \hat m_3^\dagger \hat m_3 + \hat H_{01} + \hat H_{13},
\label{hamiltonianbosons}
\end{equation}
where
\begin{equation}
\label{Feshbach}
\hat H_{01} = \alpha' \hat m_1^\dagger \hat b_A \hat b_B e^{i\omega t} + H.c.
\end{equation}
describes the hyperfine coupling of two bosonic atoms in state $\ket{0}$ to closed-channel molecules in state $|1\rangle$ with coupling strength $\alpha'$.

The second term
\begin{equation}
\label{Raman}
\hat H_{13} = \chi' \hat m_1^\dagger \hat m_3 \hat a + H.c.,
\end{equation}
describes the Raman transfer of the molecules to a deeply-bound
final state, together with the emission of a photon into the optical resonator,
with coupling strength $\chi'$.
In these equations, $\hat b_A$, $\hat b_B$ are
annihilation operators for atoms of species $A$ and $B$, $\hat a$
is the annihilation operator for photons in the single
cavity mode, and $\hat m_1$ and $\hat m_3$ are
annihilation operators for molecules in the states $\ket{1}$ and
$\ket{3}$, respectively. All the field operators satisfy bosonic
commutation relations. In Eq.~(\ref{Raman}) the field that
induces a virtual transition from the molecular state $\ket{1}$ to
a virtually excited state $|2\rangle$ is treated classically.

The dynamics of the cavity field, damped at a rate $\kappa$, is
described by the familiar master equation for a damped harmonic
oscillator \cite{Carmichael99}. In the bad cavity limit, $\kappa
\gg |\delta|,|\chi'|\sqrt{N}$, with $N$ the maximum number of
molecules, it can be adiabatically eliminated \cite{Wiseman93} to
give a master equation for the reduced density operator of the
atom-molecule system, $\hat \rho=Tr_{\rm cavity}[\hat w]$.
In this limit the two-photon detuning $\delta$ does not influence
the dynamics and we set it to zero for the rest of this paper.
This results in the master equation 
\begin{eqnarray}
\label{bose_master}
\frac{d\hat \rho}{d t} &=& -i \left[\alpha' \hat m_1^\dagger \hat b_A \hat b_B
e^{i\omega t} + H.c., \hat \rho \right] \\
&+& \gamma \left(\hat m_1 \hat m_3^\dagger \hat \rho \hat m_3 \hat m_1^\dagger
- \hat m_3 \hat m_1^\dagger \hat m_1 \hat m_3^\dagger \hat \rho + H.c. \right),
\nonumber
\end{eqnarray}
where $\gamma=|\chi'|^ 2/ \kappa$. The dissipative part is of the Lindblad form
$id\hat \rho/dt\propto \hat A^\dagger \hat \rho \hat A - \hat A^\dagger \hat A \hat \rho +H.C.$ with $\hat A = \hat m_1\hat m_3^\dagger$.

The first term on the right hand side of Eq.~(\ref{bose_master})
represents the conversion of pairs of atoms in state $\ket{0}$
into molecules in state $\ket{1}$, and the second
term leads to the amplification of molecules in state $\ket{3}$.
Note that the rapid decay of photons from the cavity prevents
their reabsorption in a $|3\rangle \rightarrow |1\rangle$
transition, resulting in the irreversible transfer of molecules to
a deeply-bound state.

\subsection{Molecule dynamics}

Typical experiments start out with all particles in the atomic
condensates and no molecules, corresponding to an initial Fock
state with $N_A$ atoms of species $A$ and $N_B$ atoms of species
$B$.
In the absence of losses the master equation (\ref{bose_master})
conserves the total particle number,
\begin{equation}
\label{number_conservation}
\frac{d}{dt} \left(n_A + n_B + 2n_1 + 2n_3 \right)=0,
\end{equation}
where $n_A$ and $n_B$ are the numbers of atoms of species $A$ and
$B$, and $n_1$ and $n_3$ are the numbers of molecules in states
$\ket{1}$ and $\ket{3}$, respectively. The evolution of the system
can therefore be described on the basis
\begin{equation}
\label{bose_basis}
\left| n,x \right\rangle := \left| n_{A/B} = N_{A/B}-n, n_1 = n-x, n_3 = x \right\rangle,
\end{equation}
where $0 \le n \le \min\{N_A,N_B\}$ is the total number of
molecules and $0 \le x \le n$ is the number of molecules in the
final state $\ket{3}$.

From the master equation (\ref{bose_master}) we can derive
equations of motion for the occupation numbers $n_i = \langle n_i
\rangle$. As usual for nonlinear systems, these lowest moments are
coupled to a hierarchy of equations for moments of growing order.
Truncating this hierarchy by factorizing higher order correlation
functions we find
\begin{eqnarray}
\label{bose_mean}
\dot{n}_3 &=& 2(1+n_3) n_1, \nonumber\\
\dot{n}_1 &=& -2(1+n_3) n_1 + \left(- i \bar \alpha e^{i \bar \omega\tau} P + c.c. \right), \nonumber \\
\dot{P} &=& - (1+n_3) P - i \bar \alpha e^{-i\bar\omega\tau}
[n_1 (n_A + n_B + 1) - n_A n_B],\nonumber \\
\dot{n}_A &=& \dot{n}_B = i \bar \alpha e^{i \bar \omega\tau} P + c.c.,
\end{eqnarray}
where we have introduced the atom-molecule correlation
\begin{equation}
P = \langle \hat m_1^\dagger \hat b_A \hat b_B \rangle,
\end{equation}
$\bar \alpha = \alpha' / \gamma$, $\bar \omega = \omega / \gamma$,
and the time derivative is taken with respect to the dimensionless
time $\tau = \gamma t$.

\begin{figure}
\includegraphics[width=0.9\columnwidth]{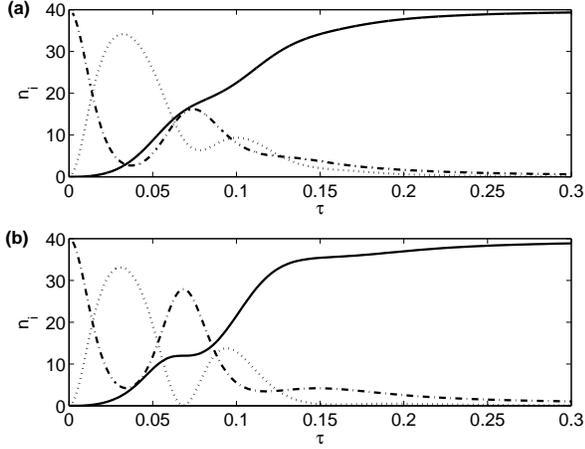}
\caption{Dynamics of molecule formation from two heteronuclear
BECs. Occupation numbers for the three modes $n_3$ (solid line),
$n_1$ (dotted line) and $n_A$ (dash-dotted line) are obtained
({\bf a}) from the master equation (\ref{bose_master}) and ({\bf
b}) from the set of mean field equations (\ref{bose_mean}). In
both cases $\bar \alpha = 10$, $\bar \omega = 0$, and
$N_A=N_B=40$.} \label{bose_fig}
\end{figure}

Fig.~\ref{bose_fig} shows the result of numerically integrating
the master equation (\ref{bose_master}) and the mean field
equations (\ref{bose_mean}) by a standard fourth-order Runge-Kutta
method for $\bar \alpha = 10$, $\bar \omega = 0$ and $N_A=N_B=40$.

The early stages of the dynamics are characterized by nonlinear
oscillations between atomic pairs in state $\ket{0}$ and the
closed-channel molecules in state $\ket{1}$ at a frequency of order
$\bar\alpha \sqrt{N_A N_B}$.
For longer times, though, the system is dominated by the two-photon
Raman transition from $\ket{1}$ to $\ket{3}$ and the damping associated
with the optical cavity losses.

While the details of the approximate results differ from the exact
dynamics, there is good qualitative agreement between
the two approaches, especially for short times. This is particularly
true for $N_A \not= N_B$, in which case an excess of one bosonic
species is always present. This leads to pronounced, more linear
oscillations of the population of atoms of the other species between
the states $\ket{0}$ and $\ket{1}$, as predicted by the mean field
equations (\ref{bose_mean}) and illustrated in Fig.~\ref{bose_fig} (b).

From simulations for varying particle numbers we infer that the dynamics
obtained from the factorized moment equations converge to the full quantum
dynamics. In the sense of this convergence $N_A=N_B=40$ is close to the
"large particle limit", at least as far as qualitative features are
concerned.

\subsection{Molecule number statistics}

An important feature of the effective Hamiltonian Eq.~(\ref{hamiltonianbosons}-\ref{Raman})
is that we can solve the resulting master equation (\ref{bose_master}) numerically and thus
we are able to determine the molecule number statistics, that is,
the probability $P^{(i)}_j(\tau)$ to measure $j$ molecules in
state $\ket{i}$ at dimensionless time $\tau$,
\begin{equation}
P^{(i)}_j(\tau) = Tr \left [ \ket{n_i = j} \bra{n_i = j} \hat \rho (\tau) \right].
\end{equation}

\begin{figure}
\includegraphics[width=0.9\columnwidth]{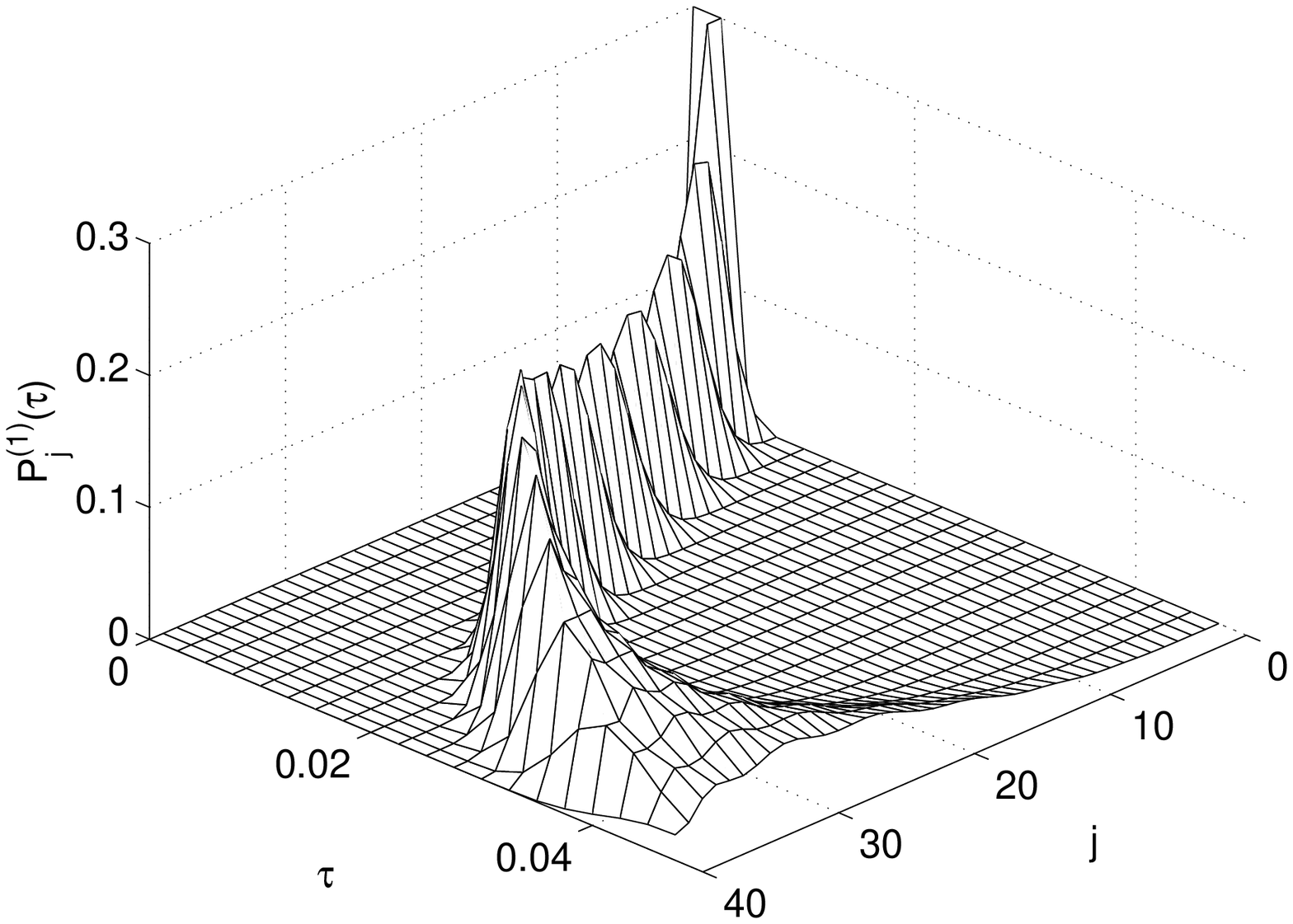}
\caption{Number statistics for the chosed-channel molecules formed
via a Feshbach resonance from a heteronuclear BEC. The parameters
are $\bar \alpha = 10$, $\bar \omega =0$, and $N_A=N_B=40$.
Probabilities are cutoff at $0.3$ for clarity.}
\label{bosestat1_fig}
\end{figure}

Fig.~\ref{bosestat1_fig} shows the time dependence of the molecule
number statistics for state $\ket{1}$, $P^{(1)}_j(\tau)$.
For short enough times this distribution is very nearly Poissonian, and
it remains so, until $\langle n_1 \rangle$ comes close to its maximum value,
i.e.~the molecular field can be considered coherent during this stage.

To support this interpretation we calculate the second-order coherence
function of the molecules in state $\ket{1}$ in the short time limit,
\begin{equation}
\label{bose_g2}
g^{(2)} = \lim_{\tau\rightarrow 0} g^{(2)}(\tau) =
\lim_{\tau\rightarrow 0} \frac{\langle \hat m_1^\dagger(0) \hat
m_1^\dagger(\tau) \hat m_1 (\tau) \hat m_1(0) \rangle}{\langle \hat
m_1^\dagger(0) \hat m_1(0) \rangle \langle \hat m_1^\dagger(\tau)
\hat m_1(\tau) \rangle},
\end{equation}
by expanding the expectation values in Eq.~
(\ref{bose_g2}) around $\tau = 0$. We find
\begin{equation}
g^{(2)} = \frac{\langle \hat b_A^\dagger \hat b_A^\dagger
\hat b_B^\dagger \hat b_B^\dagger \hat b_A \hat b_A \hat b_B
\hat b_B \rangle}{\langle \hat b_A^\dagger \hat b_B^\dagger
\hat b_A \hat b_B\rangle^2} = \frac{(N_A-1)(N_B-1)}{N_A N_B},
\end{equation}
showing that the intermediate molecules in state $\ket{1}$ are
approximately second-order coherent in the limit of large particle
numbers $N_A,N_B \rightarrow \infty$.
Moreover, we note that an excess of one species of bosonic atoms
leads to the persistence of the initial second-order coherence over
several oscillations between the states $\ket{0}$ and $\ket{1}$.

\begin{figure}
\includegraphics[width=0.9\columnwidth]{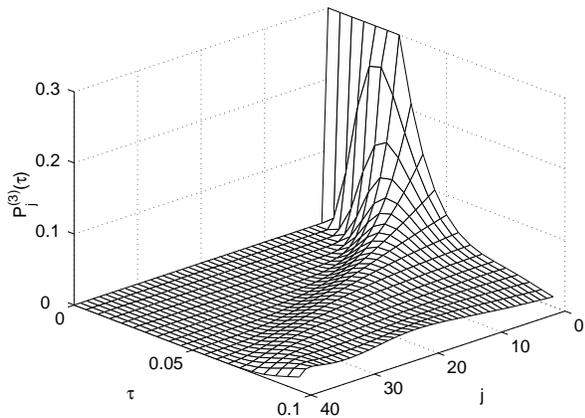}
\caption{Number statistics for ground state molecules formed from
a heteronuclear BEC. The parameters are $\bar \alpha = 10$, $\bar
\omega =0$, and $N_A=N_B=40$. Probabilities are cutoff at $0.3$
for clarity.} \label{bosestat3_fig}
\end{figure}

Fig.~\ref{bosestat3_fig} shows the molecule number statistics in
state $\ket{3}$.
For short times, the associated second-order correlation function is found to be
\begin{eqnarray}
g^{(2)} &=& \frac{\langle \hat a \hat a \hat a^\dagger \hat a^\dagger \rangle}
{\langle \hat a \hat a^\dagger \rangle}
\cdot \frac{\langle \hat b_A^\dagger \hat b_A^\dagger
\hat b_B^\dagger \hat b_B^\dagger \hat b_A \hat b_A
\hat b_B \hat b_B \rangle}{\langle \hat b_A^\dagger
\hat b_B^\dagger \hat b_A \hat b_B\rangle^2} \nonumber \\
&=& 2 \cdot \frac{(N_A-1)(N_B-1)}{N_A N_B}, \label{g23}
\end{eqnarray}
and exhibits the factor of 2 typical of a thermal field.
The antinormally ordered correlation function of the cavity mode
that appears in the first equality in Eq.~(\ref{g23}) clearly
shows that the loss of coherence suffered by the molecular field
during transfer from the intermediate state $|1\rangle$ to the
deeply-bound state $|3\rangle$ finds its origin in the vacuum
field of the optical cavity. The two-photon Raman transition can
be viewed as four-wave-mixing between two molecular fields and
two optical fields where one optical mode is in its vacuum and
thus the enhanced noise in the final molecular state is due to
the spontaneous nature of the emission of the photon:
The optical resonator acts as an input port that injects quantum noise into the
molecular system, similarly to the shot-noise injected through the empty input
port of an optical beam splitter. The enhanced noise in field $\ket{3}$ also
manifests itself in the very broad number distribution in Fig.~\ref{bosestat3_fig}.

\section{Bose-Fermi mixture}

\subsection{Model}

We now turn to the case of Feshbach-assisted photoassociation from
a Bose-Fermi mixture that consists of a BEC and a degenerate Fermi
gas. We show that this system, in a degenerate model, can be
mapped onto the case of heteronuclear BECs considered in the
previous section.

As before, the atomic condensate is treated as a zero-temperature condensate described by a single-mode bosonic field with zero momentum, background scattering as well as three-body losses are neglected.
The Fermi gas is assumed to be at $T=0$.

Differences in kinetic energy eventually lead to dephasing between the fermions. We focus on the short time behavior (shorter than $h/E_F$) where the entire Fermi gas resonds cooperatively. This leads to a degenerate description of the atoms, or homogeneously broadened in the language of quantum optics \cite{Moore01, Ketterle01}.

The hyperfine coupling results, as before, in the formation of
closed-channel molecules in state $\ket{1}$, which can then be transferred
to a deeply-bound state via a two-photon Raman process that
changes the momenta of the molecules by the difference of the
momenta of the absorbed and the emitted photons, $\vec{q}$.

In these approximations the evolution of the system is governed by
the effective Hamiltonian
\begin{equation}
\hat H = \delta \sum_{\vec k} \hat m_{3\vec k}^\dagger \hat m_{3\vec k}
+ \hat H_{01} + \hat H_{13},
\end{equation}
where
\begin{equation}
\hat H_{01}=\alpha' \sum_{\vec k} \hat m_{1\vec k}^\dagger
\hat f_{\vec k} \hat b e^{i\omega t} + H.c.
\label{bosefermi1}
\end{equation}
and
\begin{equation}
\hat H_{13}= \chi' \sum_{\vec k} \hat m_{1\vec k}^\dagger
\hat m_{3\vec k + \vec q} \hat a + H.c.
\label{bosefermi2}
\end{equation}
Here $\hat b$ is the annihilation operator for atoms in the
bosonic condensate, $\hat f_{\vec k}$ is the fermionic
annihilation operator for atoms with momentum $\vec k$, $\hat a$
the annihilation operator for cavity photons, and $\hat m_{1\vec
k}$ and $\hat m_{3\vec k}$ are the annihilation operators for
fermionic molecules in the electro-vibrational states $\ket{1}$
and $\ket{3}$ with momentum $\vec k$.
The parameters $\delta$, $\omega$, $\alpha'$ and $\chi'$ are the same as before.

We proceed by introducing the pseudo-spin operators
\begin{eqnarray}
\hat s_{\vec k}^+ &=& \left(\hat s_{\vec k}^-\right)^\dagger
= \hat m_{1\vec k}^+ \hat f_{\vec k},  \nonumber \\
\hat s_{\vec k}^z &=& \frac{1}{2} \left( \hat m_{1\vec k}^\dagger
\hat m_{1\vec k} - \hat f_{\vec k}^\dagger \hat f_{\vec k}\right), \nonumber \\
\hat t_{\vec k}^+ &=& \left(\hat t_{\vec k}^-\right)^\dagger
= \hat m_{1\vec k}^\dagger \hat m_{3\vec k + \vec q}, \nonumber \\
\hat t_{\vec k}^z &=& \frac{1}{2} \left( \hat m_{1\vec k}^\dagger
\hat m_{1\vec k} - \hat m_{3\vec k + \vec q}^\dagger \hat m_{3\vec
k + \vec q}\right),
\end{eqnarray}
as well as the total spin operators
\begin{eqnarray}
\hat S^\pm &=& \sum_{\vec k} \hat s_{\vec k}^\pm, \qquad
\hat S^z = \sum_{\vec k} \hat s_{\vec k}^z, \nonumber \\
\hat T^\pm &=& \sum_{\vec k} \hat t_{\vec k}^\pm, \qquad
\hat T^z = \sum_{\vec k} \hat t_{\vec k}^z,
\end{eqnarray}
in terms of which the effective Hamiltonians (\ref{bosefermi1}) and (\ref{bosefermi2}) become
\begin{eqnarray}
\hat H_{01} & = & \alpha' \hat S^+ \hat b e^{i\omega t} + H.c., \nonumber \\
\hat H_{13} & = & \chi' \hat T^+ \hat a + H.c.
\end{eqnarray}

Applying then twice the Schwinger mapping between an angular
momentum operator and the creation and annihilation operators of
two bosonic modes,
\begin{eqnarray}
\hat S^+ &\rightarrow& \hat m_1^\dagger \hat b_B, \nonumber \\
\hat T^+ &\rightarrow& \hat m_3^\dagger \hat m_1,
\end{eqnarray}
where $\hat b_B$, $\hat m_1$ and $\hat m_3$ are \emph{bosonic}
annihilation operators, we can finally formally map the Bose-Fermi
system under consideration onto the heteronuclear BEC of the
previous section. Note, however, that the physical interpretation
of the operators involved, and hence the physics, is different in
both cases.

\subsection{Molecule number statistics}

In the homogeneously broadened limit under consideration the
atom-molecule coupling treats all momenta identically, so that the
fermionic modes are always in a collective state, since they start
in a totally symmetric state.

Introducing the total occupation numbers of the states $\ket{1}$ and $\ket{3}$ as
\begin{equation}
\hat n_1 = \sum_{\vec k} \hat m^\dagger_{1\vec k} \hat m_{1\vec k}
\end{equation}
and
\begin{equation}
\hat n_3 = \sum_{\vec k} \hat m^\dagger_{3\vec k} \hat m_{3\vec k},
\end{equation}
the evolution of the atom-molecule system can be described in the
same basis as in Eq.~(\ref{bose_basis})
\begin{equation}
\label{bosefermi_basis}
\ket{n,x} = \ket{n_{b/f} = N_{b/f}-n, n_1 = n-x, n_3 = x},
\end{equation}
where, as before, $0 \le n \le \min\{N_b,N_f\}$ and $0 \le x \le
n$, $N_b$ and $N_f$ being the initial number of bosons and
fermions, respectively. It follows from the Schwinger mapping that
the matrix elements are the same for the Bose-Fermi mixture as for
the heteronuclear BEC.

Since all fermionic modes are treated on the same footing, the
molecule number statistics for a \emph{single} fermionic mode of
momentum $\vec k$ are simply given by
\begin{equation}
P(n_{i\vec k} = 1) = 1 - P(n_{i\vec k} = 0) = \frac{\langle n_i\rangle}{N_f},
\end{equation}
and the molecule dynamics of a \emph{single} fermionic mode
$\ket{i}_{\vec k}$ are given by Fig.~\ref{bose_fig} with the
vertical axes normalized to unity.

The statistics of the total molecule number in
states $\ket{1}$ and $\ket{3}$ are the same as their heteronuclear
BEC counterparts and are shown in Fig.~\ref{bosestat1_fig} and
Fig.~\ref{bosestat3_fig}. For short enough times the excitation of
the various momentum states of the closed-channel molecules are
independent and the number statistics in state $\ket{1}$ is
therefore
\begin{equation}
P^{(1)}_j = \binom{N_f}{j} \left[ P(n_{1\vec k} = 1)\right] ^j
\left[ P(n_{1\vec k} = 0) \right]^{N_f-j},
\end{equation}
a result valid as long as the molecule statistics for different
$\vec k$ are statistically independent. For large particle number
$N_f$, the binomial distribution converges to a Gaussian with mean
$\langle n_1 \rangle$ and variance
$\langle n_1 \rangle (1-\langle n_1\rangle / N_f )$,
very similar to a coherent state, see
Fig.~\ref{bosestat1_fig}. For longer times the various momentum
states cease to be statistically independent.

\section{Two degenerate Fermi gases}

\subsection{Model}

We finally turn to the case of Feshbach-assisted photoassociation
from two normal Fermi gases of atoms of species $A$ and $B$.
Treating the Fermi gases in the homogeneously broadened limit and
the bosonic molecules in a single-mode approximation as in section
\ref{twobecs},
the corresponding effective Hamiltonian is
\begin{equation}
\hat H = \delta \hat m_3^\dagger \hat m_3 + \hat H_{01} + \hat H_{13},
\end{equation}
with
\begin{equation}
\hat H_{01} = \alpha' \hat m_1^\dagger \sum_{\vec k}
\hat f_{A,\vec k} \hat f_{B, -\vec k}  e^{i\omega t} + H.c.
\end{equation}
and
\begin{equation}
\hat H_{13} = \chi' \hat m_1^\dagger \hat m_3 \hat a + H.c.
\end{equation}
Here $\hat f_{A,\vec k}$ and $\hat f_{B,\vec k}$ are the fermionic
annihilation operators for atoms of species $A$ and $B$ with
momentum $\vec k$, $\hat a$ the bosonic annihilation operator for
the single cavity mode, $\hat m_1$ and $\hat m_3$ the bosonic
annihilation operators for molecules in the states $\ket{1}$ and
$\ket{3}$.
The parameters $\delta$, $\omega$, $\alpha'$ and $\chi'$ are the same as before.

Introducing the pseudo-spin operators \cite{Anderson58}
\begin{eqnarray}
\label{fermi_spin}
\hat s_{\vec k}^+ &=& \left(\hat s_{\vec k}^-\right)^\dagger =
\hat f_{A,\vec k}^\dagger \hat f_{B,-\vec k}^\dagger, \nonumber \\
\hat s_{\vec k}^z &=& \frac{1}{2} \left( \hat f_{A,\vec k}^\dagger
\hat f_{A,\vec k} + \hat f_{B,-\vec k}^\dagger \hat f_{B,-\vec k} - 1 \right),
\end{eqnarray}
and the total spin operators
\begin{equation}
\hat S^\pm = \sum_{\vec k} \hat s_{\vec k}^\pm, \qquad
\hat S^z = \sum_{\vec k} \hat s_{\vec k}^z,
\end{equation}
results in the master equation for the reduced density operator
\begin{eqnarray}
\label{fermi_master}
\frac{d\hat \rho}{dt} &=& -i \left[\alpha'
\hat m_1^\dagger \hat S^- e^{i\omega t} + H.c. ,\hat \rho \right]\\
&+& \gamma \left(\hat m_1 \hat m_3^\dagger \hat \rho \hat m_3
\hat m_1^\dagger - \hat m_3 \hat m_1^\dagger \hat m_1 \hat m_3^\dagger
\hat \rho + H.c. \right). \nonumber
\end{eqnarray}

\subsection{Molecule dynamics}

Using the basis defined in Eqs. (\ref{bose_basis}) and
(\ref{bosefermi_basis}), the action of the operators $\hat S^-$ on
a state $\ket{n,x}$ is
\begin{eqnarray}
\hat m^\dagger_1 \hat S^- \ket{n,x} &=&
\hat m^\dagger_1 \hat S^- \ket{S=N/2, m_S = N/2-n}\\
&=& \sqrt{n-x+1} \sqrt{(n+1)(N-n)} \ket{n+1,x}, \nonumber
\end{eqnarray}
where $N = \min \{ N_A, N_B \}$ is the maximum number of
molecules, and $N_A$ and $N_B$ are the initial numbers of atoms of
species $A$ and $B$.

A semi-classical approximation can again be obtained by deriving
equations of motion for the expectation values of the occupation
numbers and factorizing higher-order correlations. This gives
\begin{eqnarray}
\label{fermi_mean}
&&\dot{n}_3 = 2 (1+n_3)n_1, \nonumber \\
&&\dot{n}_1 = -2 (1+n_3)n_1 + \left(- i \bar \alpha
e^{i\bar\omega\tau} P + c.c.\right), \nonumber \\
&&\dot{P}=- (1+n_3)P + i \bar \alpha e^{i\bar\omega\tau}
\left(\langle \hat S^+ \hat S^- \rangle + 2 n_1
\langle \hat S^z \rangle \right), \nonumber \\
&&\frac{d}{d\tau}\langle \hat S^+ {\hat S}^- \rangle= -2 i
\bar \alpha e^{-i\bar\omega\tau} P \langle \hat S^z - 1\rangle + c.c.,\nonumber \\
&&\frac{d}{d \tau}\langle {\hat S}^z \rangle= i \bar\alpha
e^{i\bar\omega\tau} P + c.c.,
\end{eqnarray}
where we have introduced the atom-molecule correlation
\begin{equation}
\label{P}
P = \langle \hat m_1^\dagger \hat S^- \rangle.
\end{equation}

\begin{figure}
\includegraphics[width=0.9\columnwidth]{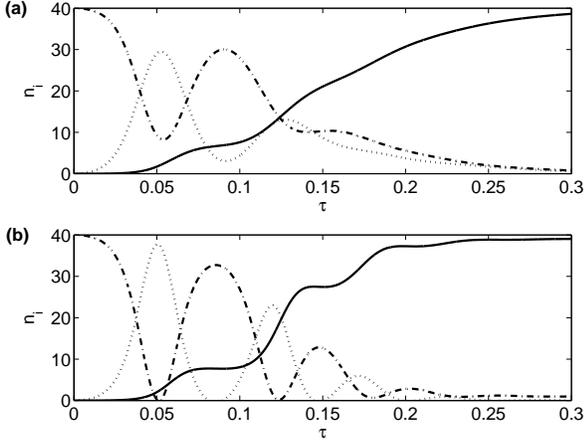}
\caption{Dynamics of molecule formation from a heteronuclear Fermi
gas mixture. Occupation numbers for the three modes $n_3$ (solid
line), $n_1$ (dotted line) and $n_A$ (dash-dotted line) are ({\bf
a}) obtained from the master equation (\ref{fermi_master}) and
({\bf b}) from the set of mean field equations (\ref{fermi_mean}).
In both cases $\bar \alpha = 10$, $\bar \omega = 0$, and
$N_A=N_B=40$.} \label{fermi_fig}
\end{figure}

Fig.~\ref{fermi_fig} shows the evolution of the atom number and the populations
of the two bosonic states as obtained from a numerical solution of the master
equation (\ref{fermi_master}) and the approximate semiclassical
equations of motion (\ref{fermi_mean}) for $\bar \alpha = 10$,
$\bar \omega = 0$, and $N_A=N_B=40$.
We again observe nonlinear atom-molecule oscillations between
$\ket{0}$ and $\ket{1}$.
The mean-field equations still describe the qualitative features
of the full dynamics, although the agreement is not as good as
in the bosonic case
\footnote{Note that these equations lead to the occasional
appearance of unphysical negative occupation numbers, a well known
effect arising from the inclusion of non-gaussian fluctuations such
as $P$ in Eq.~(\ref{P}).}.
Note that the slope of $\langle n_1(\tau)\rangle$ near $\tau=0$ is different
from the bosonic case above. This finds its explanation in the different
molecule statstics to be discussed next.

\subsection{Molecule number statistics}

\begin{figure}
\includegraphics[width=0.9\columnwidth]{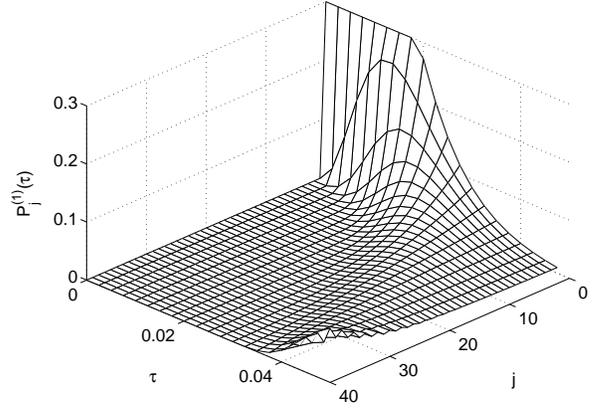}
\caption{Number statistics for the closed-channel molecules formed
via a Feshbach resonance from a heteronuclear Fermi gas mixture.
The parameters are $\bar \alpha = 10$, $\bar \omega =0$, and
$N_A=N_B=40$. Probabilities are cutoff at $0.3$ for clarity.}
\label{fermistat1_fig}
\end{figure}

Fig.~\ref{fermistat1_fig} shows the molecule statistics for state $\ket{1}$.
In contrast to bosonic atoms, they are now reminiscent of a
thermal state, a result corroborated by the second-order coherence
function
\begin{equation}
g^{(2)} = \frac{\langle \hat S^+ \hat S^+ \hat S^- \hat S^- \rangle}
{\langle \hat S^+ \hat S^- \rangle^2} = \frac{2(N-1)}{N},
\end{equation}
characteristic for a chaotic light field. This is because the
fermionic atoms act as independent "radiators", much like the
atoms in a thermal light source, leading to the different slope of $\langle
n_1(\tau)\rangle$ near $\tau=0$ in the bosonic and fermionic case mentioned
above.

\begin{figure}
\includegraphics[width=0.9\columnwidth]{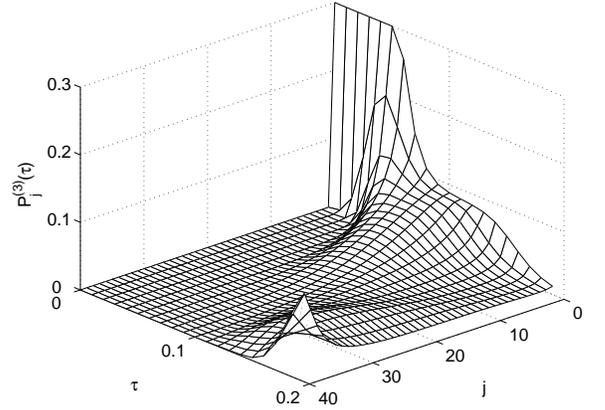}
\caption{Number statistics for ground state molecules formed from
a heteronuclear Fermi gas mixture. The parameters are $\bar \alpha
= 10$, $\bar \omega =0$, and $N_A=N_B=40$. Probabilities are
cutoff at $0.3$ for clarity.} \label{fermistat3_fig}
\end{figure}

Fig.~\ref{fermistat3_fig} shows the molecule number statistics in
the deeply-bound state $\ket{3}$.
As already discussed, the coupling of the molecular field to the
lossy optical cavity provides an input port for quantum noise, so
that the number fluctuations in that state are twice those in the
intermediate state $|1\rangle$. Specifically, we find
\begin{equation}
\label{fermig2m3}
g^{(2)} = \frac{\langle \hat a \hat a \hat a^\dagger \hat a^\dagger\rangle}
{\langle \hat a \hat a^\dagger\rangle} \cdot \frac{\langle \hat S^
+ \hat S^+ \hat S^- \hat S^- \rangle}{\langle \hat S^
+ \hat S^- \rangle^2} = 2 \cdot \frac{2(N-1)}{N},
\end{equation}
which leads for $N\rightarrow \infty$ and short times to $g^{(2)}=4$.

We conclude this section by considering two Fermi
gases subject to an attractive interaction. The ground state of
such a system is approximately given by the BCS state
\begin{equation}
\ket{BCS} = \prod_{\vec k} (u_{\vec k} + v_{\vec k} s_{\vec k}^+) \ket{0},
\end{equation}
with Bogoliubov amplitudes $u_{\vec k}$ and $v_{\vec k}$ and the
atomic vacuum state $\ket{0}$.

The second-order coherence $g^{(2)}$ of closed-channel molecules in
state $\ket{1}$ becomes in that case
\begin{equation}
g^{(2)} = 2 - \frac{(\Delta/V)^4}{\left(\langle N \rangle + (\Delta/V)^2\right)^2},
\end{equation}
where $\Delta = V \sum_k u_k v_k$ is the gap parameter, $V$ is the
attractive two-body potential between fermionic atoms, and
$\langle N \rangle = \sum_k |v_{\vec k}|^2$ the average particle
number in the system.

For state $\ket{3}$ we get
\begin{equation}
g^{(2)} = 4 - \frac{2(\Delta/V)^4}{\left(\langle N \rangle + (\Delta/V)^2\right)^2},
\end{equation}
which again differs from the result for state $\ket{1}$ by the
cavity factor
$\langle \hat a \hat a \hat a^\dagger \hat a^\dagger\rangle / \langle \hat a \hat a^\dagger \rangle= 2$.

For $\Delta = 0$ we recover the results for a normal Fermi gas
$g^{(2)} = 2$ for state $\ket{1}$ and $g^{(2)} = 4$ for state
$\ket{3}$. As the number of Cooper pairs $\Delta/V$ becomes
macroscopic, which is the case for a superfluid system, we recover
the results for bosonic atoms $g^{(2)} = 1$ for state $\ket{1}$
and $g^{(2)} = 2$ for state $\ket{3}$. This demonstrates that
superfluidity results in a degree of coherence of the molecules intermediate
between those of a BEC and of a normal Fermi gas.

\section{Discussion}

In the single mode approximation for bosons and
degeneracy approximations for fermions, the dynamics of Feshbach-assisted
photoassociation of ultracold molecules from atoms with different statistics
can be treated in a unified framework. At the level of molecular populations,
the molecule dynamics are to a large extent independent of the quantum
statistics of the atoms. In the limit of homogeneously broadened fermionic
systems, the descriptions of molecule formation from bosonic atoms
and from Bose-Fermi mixtures can be mapped onto the same
Hamiltonian with identical initial states. Although the
interpretation of the results in both cases is different, with
obvious fundamental differences at the detailed level of
description of the molecular fields, their global properties, when
summed over the momenta of the molecules, are therefore largely
similar \cite{Wouters03,Dannenberg03}.

The similarities in the generation of closed-channel molecules from
bosons and fermions stem from the collective behavior of these
systems, which are well known e.g.~in the context of matter-wave
four-wave mixing \cite{Moore01,Ketterle01}. Bose enhancement is
built into the state symmetrization procedure for bosons, while
the collective behavior of fermions results from the constructive
interferences that occur when a collective state is built up by an
interaction that cannot distinguish between the different particles.
In addition, the two-photon Raman process
that transfers molecules from the intermediate state $\ket{1}$ to
their final state $\ket{3}$ does not distinguish whether the
bosonic molecules are built out of bosons or fermions.

Despite this analogy, though, the statistics of the resulting
molecular fields exhibit a distinct signature of the type of atoms
from which they are formed. This difference stems from the fact
that in case the molecules are formed from pairs of atoms in Fermi
gases, these atoms act initially as independent "radiators", much
like independently radiating atoms in a chaotic light source,
whereas molecule formation from atomic Bose-Einstein condensates
is a coherent process from the very beginning \cite{Meiser05}.

In the degenerate model the kinetic energies of fermions are neglected. They destroy the collective effects that give rise to the similarities between bosons and fermions at times larger than $h/E_F$ \cite{Moore01, Ketterle01}.
Uys {\it et al.}~\cite{Uys05} compared molecule association in a degenerate to a non-degenerate model and found the short time behavior indeed indistinguishable between the two, whereas the overall conversion efficiency is significantly reduced.
Szymanska {\it et al.}~\cite{Szymanska05} studied the short-time dynamics following an abrupt jump in the atomic interaction for the entrance channel dominated resonances observed in $^{40}K$ and $^6Li$. They point out that in these systems large-scale atom-molecule oscillations are precluded. We stress that, in contrast to \cite{Szymanska05}, our studies focus on Feshbach resonances which are closed channel dominated.
Nevertheless, for longer times, the dephasing between the fermions will destroy the collective effects and reduce the conversion efficiency.

Atom-atom, atom-molecule and molecule-molecule collissions lead to several modifications of the theory as it has been presented here.
Interactions between atoms in a BEC give rise to quantum depletion, that can be taken into account by means of the Bogoliubov approximation for the noncondensed part. Usually, these corrections are very small with the dominant physics well captured by the single mode approximation we use in this paper. For fermions repulsive interactions lead to very minor quantitative changes that are hard to distinguish from excitations due to finite temperature. Attractive interactions between fermionic species on the other hand give rise to the BCS phase transition and we have included them in the framework of the BCS mean field theory.
Atom-molecule interactions and interactions between atoms of the two different atomic species may lead to phase seperation \cite{Molmer98}. For closed-channel dominated resonances however the background scattering is usually neglectable if the system is tuned close to the resonance \cite{Wouters03}.
Inelastic collisions give rise to losses which can easily be incorporated in our formalism through a phenomenological damping term, e.g.~in Eq.~(\ref{bose_mean}). Calculations of the full counting statistics, however, become much harder numerically, as Eq.~(\ref{number_conservation}) ceases to hold. It is worth noting that losses have no effect on our conclusions which are concerned with the short time properties of the molecule formation process. Thus, we are justified in omitting them.

The second-order coherence properties of the final ground state
molecules are further influenced by their coupling to a lossy
optical cavity that acts as an input port for quantum noise, in a
fashion reminiscent of the role of the empty input port in an
optical beam splitter. This process, which is dominated by vacuum
fluctuations, results in an additional factor of two enhancement
in the second order coherence $g^{(2)}$, both for bosonic and
fermionic gases.

Future work will extend our model to take into account non-condensed modes of the bosonic fields. This will allow us to study finite temperature effects as well as dynamical depletion of the condensate in the course of the molecule formation. Furthermore, we will use a more detailed description of the two-body physics. In doing so we hope to be able to study the effects of a time-varying detuning, which is necessary to adequately describe sweep experiments.

\acknowledgments
We thank Keith Burnett, Krzysztof G\'oral, and Thorsten K\"ohler for helpful
discussions.  This work is supported in part by the US Office of Naval
Research, by the National Science Foundation, by the US Army Research Office,
and by the National Aeronautics and Space Administration. A.N. gratefully
recognizes a scholarship from the Studienstiftung des deutschen Volkes.

\end{document}